# Universal coherent structures of elastic turbulence in straight channel with viscoelastic fluid flow

Narsing K. Jha[1] and Victor Steinberg[1,2*]

[1]Department of Physics of Complex Systems, Weizmann Institute of Science, Rehovot 76100, Israel

[2]The Racah Institute of Physics, Hebrew University of Jerusalem, Jerusalem 91904, Israel.

*Corresponding author. Email: victor.steinberg@weizmann.ac.il

**Abstract:**

In the present study, we investigated flow structures and properties of elastic turbulence in straight 2D channel viscoelastic fluid flow and tested earlier observations. We discovered self-organized cycling process of weakly unstable coherent structures (CSs) of co-existing streaks and stream-wise vortices, with the former being destroyed by Kelvin-Helmholtz-like instability resulting in chaotic structures. The sequence periodically repeats itself leading to stochastically steady state. This self-sustained process (SSP) remarkably resembles one investigated theoretically and experimentally for Newtonian turbulence in straight channel flow. The unexpected new ingredient is the observation of elastic waves, which finds to be critical for existence of CSs and SSP generation due to energy pumping from large to smaller scales preceding sharp power-law decay in elastic turbulence energy spectrum. The reported finding suggests the universality of CSs in transition to turbulence via self-organized cycling (SSP) in linearly stable plane shear flows of both elastic and Newtonian fluids.

**One Sentence Summary:** Universality of CSs in transition to turbulence via self-organized cycling (SSP) in linearly stable plane shear flows of both elastic and Newtonian fluids.

## Introduction

During the last about thirty years, there has been a significant progress in theoretical and experimental understanding of transition to turbulence via finite-size perturbations of linearly stable plane shear flows of Newtonian fluid. It is based, on the development of a non-modal (or non-normal) algebraic growth of perturbations *(1)*. This development has been preceded by a mathematical formulation and elaboration of a new conception of the non-modal instability instead of widely accepted and used the least stable normal eigen-mode exponential growth of infinitely small perturbations *(2)*. The former may bypass the latter on short-time scales and reach on intermediate time scales, where the large perturbation amplitudes are sufficient for significant nonlinear interactions resulting in coherent structures (CSs) and self-sustained cycling process (SSP) in transition to turbulence at large Reynolds numbers, Re>>1 *(1)*.

In the present study, we deal with a viscoelastic fluid flow, and specifically a dilute polymer solution flow. A tiny addition of long, flexible polymer molecules strongly affects laminar as well as turbulent flows, and for the former resulting in a chaotic flow called elastic turbulence (ET). ET is attributed to polymer stretching, which generates elastic stress and its back reaction on the flow. Its properties are analogous to those observed in hydrodynamic turbulence, although the formal similarity does not imply a similarity in physical mechanisms underlining these two of random motion. ET is observed at Re<<1 and the Weissenberg number, Wi>>1, the second control parameter in the problem *(3-5)*. Though instabilities and ET are observed in flows with curvilinear streamlines, they vanish in the limit of zero curvature *(3,6)*. Here, both Re=$\rho UL/\eta$ and Wi=$\lambda U/L$ are defined via the mean fluid velocity U and the vessel size L, $\rho$, $\eta$ and $\lambda$ are the density, dynamic viscosity and longest polymer relaxation time of the fluid, respectively. Thus, ET is a chaotic, inertialess flow driven solely by nonlinear elastic stress generated by polymers stretched by the flow, which is strongly modified by a feedback reaction of the elastic stress *(7)*.

Even though a linear stability analysis of elastic parallel shear flows predicts their absolute stability at all Wi *(3)*, recent theoretical developments *(8-12)* suggest two different mechanisms of possible transition to ET in these flows. Recent experiments *(13-16)* reveal large velocity fluctuations resulting from large external perturbations. Moreover, authors of *(16)* claim observing a hysteresis predicted in *(8)*, whereas predicted traveling waves are not reported. It suggests that an elastic instability and probably ET may occur in parallel shear flows. The linear stability of elastic parallel shear flows does not preclude the possibility of a transition to ET via a scenario similar to that discovered and shortly discussed above in parallel shear Newtonian fluid flows *(1,17)*. Two different theoretical approaches to this problem in viscoelastic shear flows pointed out above are undertaken. First is to use subcritical finite-amplitude solutions of the perturbation amplitude equation with a sufficient number of terms in a series of the elastic stress complex amplitude. The amplitude equation is derived from the basic equations for plane Couette flow (PCF) of the Upper-Convected-Maxwell polymer model *(18)* at Re<<1 and Wi>Wi$_c$, where Wi$_c$ is the transition value. Then a hysteretic, subcritical bifurcation results in 2D *stable* traveling waves later on leading to ET *(8)*. The second approach utilizes recently developed theoretically and verified experimentally non-modal analysis *(2)* of linearly stable shear flows of Newtonian fluid at Re>>1 *(1,2,19)*, which reveals a transient energy amplification, to explain instability of linearly stable both PCF and plane Poiseuille flow (PPF) of viscoelastic fluid at Wi>>1 *(9-12)*. Then the first transition to ET in elastic parallel shear flows is also predicted to occur via weakly unstable stream-wise vortices *(11)* or streaks *(12)*, which are the most amplified CSs, similar to those observed in Newtonian PCF at Re>>1 *(19)*. In the Newtonian flows, the regeneration of CSs occurs via dynamics described by self-sustained process (SSP), which is identified theoretically and verified experimentally *(20)* during the last two decades *(1,17,19)*. Nevertheless, the key difference between inertial and elastic instabilities is in different fields responsible for instability and so perturbations of these fields: velocity in inertial turbulence versus elastic in ET, and the latter is trickier to excite. In this paper, we present experimental observations of CSs and SSP scenario in a straight channel of viscoelastic flow in ET at Re<<1 and Wi>>1.

## Results

The schematic of the experimental setup is shown in Fig. S1 *(Suppl. Mater.)*, whereas measurement techniques, solution preparation and characterization of polymer solution are presented in *(Materials and Methods)*. The flow is globally characterized by the normalized friction factor $C_f$ /$C_f^{lam}$ versus Wi for wide range of about 1 to 500 corresponding to Re$\approx$0.001-0.35 for three normalized distances from the obstacle array l/h, where $C_f$=2H$\Delta p$/$\rho U^2\Delta L$, H=2wh/(w+h)=0.875 mm, $\Delta p$ is the pressure drop over $\Delta L$, and U=<Q>/$\rho$wh measured via the average flow discharge rate <Q>=<$\Delta$m/$\Delta$t> *(21,22)*. For l/h of 26-82, the dependence of $C_f$ /$C_f^{lam}$ on Wi reveals three regimes (Fig. 1A): transition with $C_f$/$C_f^{lam}$ ~$Wi^\gamma$ and $\gamma\approx$0.125, ET with slight $C_f$/$C_f^{lam}$ reduction and a large scatter, and drag reduction (DR) with $C_f$/$C_f^{lam}$ ~$Wi^{-\delta}$ and $\delta\approx$0.2. For l/h from -30 to 16 across cylinders (inset in Fig. 1A), the exponents are similar to those found in *(21)*. At l/h=200-250, large scatter around $C_f$ /$C_f^{lam}$ $\approx$1 is detected (Fig. 1A, open circle) and $C_f$/$C_f^{lam}$ is of around 1 for Newtonian solvent. Growth of $C_f$/$C_f^{lam}$ is associated with a sharp increase of stream- and span-wise velocity as well as pressure fluctuations (Fig. 1B, C, D, E). One notices that the scaling exponents differ significantly from that found in *(16)*, where DR is also not detected.

The key finding of the present study is CSs and SSP presented in the range of l/h=36-41 in ET at Wi=185 and at four moments (t*=tf$_{el}$) in 4 rows, where f$_{el}$ is the elastic wave frequency *(22)*. In Fig.

2 (and Movie S1), three columns show: (i) stream-wise velocity fluctuations u', (ii) vertical vorticity fluctuations ω', and (iii) span-wise gradient of span-wise velocity fluctuations $\frac{\partial w'}{\partial z}$, non-dimensionalized using U and h/U. At t*=0.15 and 0.29 (first and second row respectively), one notices fast and slow u' streaks separated by interfaces with u'=0. At t*=0.56 (third row), the lower streaks become highly perturbed, and at t*=0.67 all streaks are destroyed, probably, due to elastically-driven Kelvin-Helmholtz-like instability. Then a new cycle starts. The second column presents stripes of positive and negative ω' associated with the streak interface fluctuations at t*=0.15, 0.29 and 0.56 (upper streak interface), whereas at t*=0.67 (fourth row), u' and ω' are random in the whole field of view. The third column shows $\frac{\partial w'}{\partial z}$, where one notices a uniform distribution in stream-wise direction representing a stream-wise roll at t*=0.15 and 0.56 (see the schematic of the detection of a stream-wise vortex from 2D PIV in a horizontal xz plane using $\frac{\partial w'}{\partial z}$ measurements in Suppl. Mater. Fig. S2 with explanation in *Suppl. Mater. Notes*), whereas at t*=0.67 a random structure develops. At the same l/h, the same analysis for different flow regimes at different Wi discloses different scenarios (see Figs. S3, S4 and S5). At Wi=55 in the transitional regime, only streaks separated by rather sharp interfaces are found, the rolls are absent, and cycling period is observed (Fig. S3 and Movie S2). At Wi=325 in the DR regime, both CSs are observed, though they are not well identified due to perturbations, and cycling exists (Fig. S4 and Movie S3). Finally at Wi=727 post DR, both CSs are well defined and the clear cycle exists (Fig. S5). Further downstream at l/h=210-217, one finds just a random flow structure without any cycling at Wi=146 (Fig. S6 A-C) and the velocity power spectrum of w' at Wi=100 and l/h=193 shows only a broad, low intensity peak of elastic waves (Fig. S6 F).

The CS dynamics are also characterized by space-time plots of u'/U and w'/U at Wi =55>Wi$_c$, Wi=185 (ET), Wi=325 (DR) at l/h=38.6 (Fig. 3A-F) and at Wi=146 and further downstream at l/h=210-217 (Fig. S6D, E), where noisy periodicity at various Wi is identified by the frequency peak in the power spectrum of w' associated with elastic waves (Fig. 4A-D). Indeed, using the same approach as in *(21,23)*, one gets the elastic waves speed c$_{el}$=Δx/τ$_p$ for each Wi, where τ$_p$ is the lag time of the peak of cross-correlation function of span-wise velocities C$_w$(Δx,τ) (Fig. 3G). The dependence c$_{el}$ ~ (Wi-Wi$_c$)$^ς$, ς=0.85±0.1 is presented in Fig.3H in accord with *(22)*. Energy of the elastic wave is characterized by the peak value in the kinetic energy spectrum in three flow regimes at l/h=38.6 and varies non-monotonically growing first from transition to ET regimes, decreases towards DR and then strongly increases at very high Wi (Fig. 5A,B). The w' energy spectra in three flow regimes are characterized by the elastic wave peak at low frequencies and the power-law decays as S$_v$(w)~ ν$^{-α}$ at higher frequencies with: α=2.5 at Wi=39, 3.2 at Wi=144, 2 at Wi=290 and 2.5 at Wi=737 (Fig. 4A), in more details power spectra at l/h=3, 15 and 193 are shown in Fig. S7. Another distinctive feature of the ET regime is an algebraic decay of pressure power spectrum with the exponent β≈ 3 in accord with *(24)* verified in other numerous flow geometries with curvilinear streamlines. The pressure power spectrum measurements are presented in three flow regimes up to Wi=400 and are independent of l/h (Fig. S8). In ET the pressure power spectrum decays with β≈ 3 in accord with the previous results *(24)*, in the transition regime-β≈ 2, and in the DR regime one finds scaling with β ≈ 3 in a shorter range of low frequencies and at higher frequencies the dependence changes.

Most of the results presented are taken at specific locations along the channel at ~30<l/h<140. It is not occasional location but defined by a propagation of perturbations generated by the cylinder array along the channel. According to the plot in Fig. 6A, an initial perturbation, which reaches about 60% for u$_{rms}$/u$_m$ at l/h=0, decays rather fast an order of magnitude down to 5-6% at l/h≈30 and then rises up to about 10% at l/h>30, a reasonable value for ET *(21,23)*, where u$_m$ is the local mean velocity. Then this turbulence intensity level persists up to l/h≈140 and reduces down to 4% at l/h≈217. This dependence of the perturbation intensity along the channel suggests that the rise of the perturbation intensity after its initial drop occurs due to an intrinsic flow instability leading to the new ET regime as also observed in the span-wise velocity power spectra in Fig. S7A. It also agrees with the observation made in *(14)*. Therewith, the ET regime does not continue until the channel outlet but ceases much earlier. Such behavior is known in a parallel shear flow of Newtonian fluid. On the other hand, a simple estimate of an advection length of a perturbation with the average flow velocity U during the polymer relaxation time $\lambda$ gives $L_{adv}=U\lambda \approx 0.1 - 1.7$ m (200-3400 l/h). It means that the perturbation advection is irrelevant to the onset of ET. Further issue is the dependence of both rms stream-and span-wise gradients of the stream-wise velocity at l/h=38.6 on Wi responsible for polymer stretching in ET. Surprising similarity of both the rms gradients presented in Fig. S10C, D indicates that elastic stress and polymer stretching are isotropic, even though the flow field is highly anisotropic. Moreover, such similarity is also found in ET at Wi=150 in the spatial dependence (Fig 6B, C) in both the gradients of stream-wise as well as span-wise velocities further away from the cylinder array in the stream-wise direction at l/h between about 30 and 140 that also indicates the isotropy of the elastic stress and polymer stretching along the channel for the highly anisotropic flow field. However, near the cylinders the rms velocity gradients of the stream-wise velocity u in x and z directions differ being higher in z direction.

**Discussion**

The main message of the paper is the discovery of SSP of the two weakly unstable most amplified CSs, namely coexisting stream-wise streaks and stream-wise vortex, which show up in three flow regimes though with different scenarios observed at l/h≈36-41. At l/h≈26-82, these regimes are characterized by different scaling exponents of $C_f/C_f^{lam}$ versus Wi (Fig. 1); the span-wise velocity as well as pressure power spectra exhibit in ET the power-law decay with the exponents $\alpha \approx 3.2$ (Figs. 4, S7) and $\beta \approx 3$ (Fig. S8), respectively, characteristic to the values found in ET in various flow geometries with curvilinear streamlines *(4,5,23,24)*. The periodic sequence of CSs repeats itself in SSP presenting a stochastically steady state that is surprisingly similar to SSP of CSs observed and thoroughly investigated in Newtonian fluid shear flows *(1,2,17,19,20)* (see also schematic of CSs cycling in Fig. S9). At l/h=210-217, only random flow structure is found.

Another key finding is the observation of the elastic waves *(22)* as a part of SSP, which plays the key role due to their remarkable contribution into both the amplification of weakly unstable CSs in viscoelastic shear flows by energy pumping at large scale and the synchronization of the cycling scenario. They strongly recall transverse shear waves (Tollmien-Schlichting waves) observed in PCF turbulence of a Newtonian fluid as an exact solution *(16)*, which take part in a cycling sequence of instabilities *(2)*. Moreover, the elastic wave frequency determines a cycling period of CSs and SSP ($f_{cycle}$) that follows from their almost identical values and dependence on Wi (Fig. 5C). The synchronization of SSP cycling and energy pumping mechanisms from large to small scales strongly depends on the intensity of the elastic waves defined by its peak value in the kinetic energy spectrum. The latter varies drastically, more than 16 times with Wi at l/h=38.6 depending on the flow regimes (Fig. 5B). The higher the intensity, the better the SSP cycle synchronization and the better the images of CSs (see Figs. 2, S3, S4, S5). Moreover, at l/h=200, only the elastic wave low intensity peak is observed in the kinetic energy spectrum without the sharp decay characteristic to ET, and a random flow structure of a velocity field (Fig. S6 A-F). At l/h=3, the elastic wave peak is absent and the algebraic decay of the velocity power spectrum shows the exponent of about -1.7 in all regimes (Fig S7B). A bit further downstream at l/h= 15, a low intensity elastic wave peak appears at higher Wi=120 but still without ET features (Fig. S7A).

Further, we extend the claim of similarity of ET in 2D viscoelastic channel with the classical turbulence picture in parallel shear flows. Based on this similarity, one may suggest the universality in the transition via the weakly unstable and most amplified CSs and the SSP cycling scenario to fully developed either inertial or elastic turbulence in such flows distinguished by their linear stability. In ET, the lowest frequencies (or the largest time and length scales) are associated with the elastic waves as seen in the span-wise velocity power spectra (Fig. 4). Moreover, as pointed out already in *(22)*, due to the appearance of the frequency peak, the algebraic decay of the kinetic energy spectrum, the distinctive feature of ET, is pushed much further up to two orders of magnitude towards the higher frequencies due to energy pumping to the smaller scales. This early suggestion is confirmed experimentally by the current data due to the observation of the decisive dependence of the onset of CSs and SSP as well as the ET energy spectrum decay on the value of the elastic wave intensity. It is clearly demonstrated that at a very short distance from the obstacle array generated perturbations in the flow (Fig. S7B, l/h=3), where the elastic wave peak in the kinetic energy spectrum still does not appear, though the perturbation intensity is the highest, the flow is random with very high velocity fluctuations. Nevertheless, ET is absent. On the other hand, further downstream from the obstacles (Fig.S7A, l/h=193), where the value of the peak intensity of the elastic waves in the kinetic energy spectrum is very low, there is also not any sign of the energy spectrum decay, characteristic to ET. It means that the energy pumping due to the elastic waves is too low and not sufficient to support CSs, SSP and ET even at high Wi. Thus, these findings strongly recall a classical turbulence picture, where energy production and pumping are sufficiently large and occur at the largest length scale, defined by a system size, towards the smallest dissipation scales via an inertial range that shows in the kinetic energy spectrum an algebraic decay characteristic for inertial turbulence *(25)*.

*Comparison with the results of others.* There has been some nice experimental observations of elastic instability in square channel flows *(13-16)*. However, the main predictions of one of two relevant theories (8) are not confirmed, in spite of the fact that the results of the experiments *(13-16)* showing large velocity fluctuations with external perturbations are attributed to the predictions of this theory. First, the authors of *(14)* claim observation of a hysteresis, as a main characteristic feature of the theory predicted in *(8)*, in the dependence of the transition to a new flow state, whereas the predicted *stable* traveling waves are not reported. Moreover, the prediction of the hysteresis in the inverse bifurcation of the order parameter is directly coupled to the presence of the saddle point *(8)*, which is absent in the experimental plot in *(14)* that looks more as a continuous transition. Second, it is emphasized in the figure caption to Fig. S1 that, in a contrast to our channel of a rectangular cross-section with the aspect ratio 7, close to that used in 2D PPF *(25)*, a channel used in *(14-16)* is of a square cross-section. So lacking of both the saddle point and traveling waves in the experimental observations, casts doubt on the claimed agreement with the theory *(8)*. On the other hand, our test for the hysteresis, presented in Fig. S10A, B, clearly indicates completely reversible paths in the dependencies of $dP_{rms}$ and $C_f/C_f^{lam}$ on Wi. On the other hand, the difference between a random structure in a space-time plot of u' on extremely large time scale *(15)* and our observations of space-time plots of both u' and w' on a minor time scale as well as CSs and SSP may be attributed to different channel geometries, though the origin of the random structure in *(15)* is not clear.

To summarize, we discovered a self-organized cycling process of weakly unstable CSs of the co-existing streaks and stream-wise vortices with the former being destroyed by Kelvin-Helmholtz-like instability resulting in chaotic structures. The sequence periodically repeats itself leading to a stochastically steady state. The unexpected new ingredient of SSP is the observation of the elastic waves, which are proved to be critical for the existence of CSs and SSP and their generation due to the elastic wave energy pumping from large to smaller scales of the sharp power-law decay in the elastic turbulence kinetic energy spectrum and also synchronize the SSP cycling. These CSs and SSP remarkably resemble those investigated theoretically and experimentally in a Newtonian turbulence of straight channel flow. Its similarity to the turbulent channel flow of Newtonian fluid suggests the universality of such a stochastically steady state. The only difference between ET and inertial turbulence is the fact that the cycling frequency $f_{cycle}$ is defined by $f_{el}$ for both the velocity and elastic stress fields in ET.

**Materials and Methods**

*Preparation and characterization of polymer solution.* As a working fluid, a dilute polymer solution of high molecular weight polyacrylamide (PAAm, $M_w = 18$ MDa; Polysciences) at concentration $c = 80$ ppm ($c/c^* \simeq 0.4$, where $c^* = 200$ ppm is the overlap concentration for the polymer used [23]) is prepared using a water-sucrose solvent with sucrose weight fraction of 64%. The solvent viscosity, $\eta_s$, at 20 °C is measured to be 130 mPa·s in a commercial rheometer (AR-1000; TA Instruments). An addition of polymer to the solvent increases the solution viscosity, $\eta$, of about 30%. The stress-relaxation method [23] is employed to obtain the longest polymer relaxation time ($\lambda$) of the solution and it yields $\lambda = 17 \pm 0.5$ s.

*Flow discharge and pressure fluctuation measurements.* The fluid exiting the channel outlet is weighed instantaneously $m(t)$ as a function of time $t$ by a PC-interfaced balance (BA210S, Sartorius) with a sampling rate of 5 Hz and a resolution of 0.1 mg. The time-averaged fluid discharge rate is $<Q> = <\Delta m/\Delta t>$. To improve the resolution and sensitivity of the friction factor measurements, pressure sensors with different resolution and pressure ranges are used. Different pressure sensors from Honeywell are used with a range of 10 inch water column, and 1, 5 and 15 PSI at a sampling rate of 1000 HZ through NI data acquisition system. It has an accuracy of 0.1% and 0.25% of full scale for HSC and SSC series, respectively. Measurements are used for friction factor and $dP_{rms}$ measurements. For absolute pressure spectra measurement, SSC series pressure sensors from Honeywell of 15 and 30 PSI range with an accuracy of 0.25% of full scale is used. For $C_f$ measurement, data is acquired for 180 sec, which is about 10 times of $\lambda$. For pressure spectra, it is acquired for 1000 sec that provides us 1 million data points for reasonable statistics. The Weissenberg and Reynolds numbers are defined as Wi=$\lambda$U/w and Re= Uh$\rho/\eta$, respectively, and U=$<Q>$/$\rho$wh with fluid density $\rho = 1310$ Kg m$^{-3}$.

*Imaging system.* For flow visualization, the solution is seeded with red fluorescent particles of diameter 3 μm (R0300B, Thermoscientific) and illuminated uniformly by 2.5 W green laser (MGL-F-CNI Laser) at 532 nm wavelength. The region is imaged in the mid-plane via a microscope objective (Leitz Wetzlar 2.5X/0.008 and 4X/0.12), and a fast camera from Photron (mini UX 100) with a spatial resolution 1.2 Mpxl at a rate of up to 3200 fps. Images are acquired with low and high spatial resolutions for temporal velocity power spectra and flow structure, respectively. Three hundred thousand velocity fields are used to obtain velocity power spectra. We perform particle image velocimetry (PIV) using PIVlab [26,27] to obtain the spatially-resolved velocity field $\mathbf{U} = (u, w)$ in the several regions at different distances from the obstacle array. The acquired images were processed using FFT based correlation with four-step box size refinement, with the final box size being 32 pixel × 32 pixel (~64 × 64 μm$^2$). Correlation box overlap was maintained at 50 % in all cases, with each box having roughly at least 4–6 particles to ensure strong correlations.

**References:**

1. F. Waleffe, On a self-sustaining process in shear flows. *Phys. Fluids* **9**, 883-900 (1997).
2. P. J. Schmid. Nonmodal stability theory. Ann. Rev. Fluid Mech. 39, 129 (2007).
3. R. G. Larson, Instabilities in viscoelastic flows. *Rheol. Acta* **31**, 213- (1992).
4. A. Groisman, V. Steinberg, Elastic turbulence in a polymer solution flow. *Nature* **405**, 53- (2000).
5. A. Groisman, V. Steinberg, Efficient mixing at low Reynolds numbers using polymer additives. *Nature* **410**, 905- (2001).

6. K. Pakdel, G. H. McKinley, Elastic instability and curved streamlines. *Phys. Rev. Lett.* **77**, 2459- (1996).

7. A. Fouxon, V. Lebedev, Spectra of turbulence in dilute polymer solutions. *Phys. Fluids* **15**, 2060-2072 (2003).

8. A. N. Morozov, W. van Saarloos, Subcritical finite-amplitude solutions for plane Couette flow of viscoelastic fluids. *Phys. Rev. Lett.* **95**, 024501 (2005).

9. N. Hoda, M. R. Jovanovich, S. Kumar, Energy amplification in channel flows of viscoelastic fluids. *J. Fluid Mech.* **601**, 407- (2008).

10. M. R. Jovanovich, S. Kumar, Nonmodal amplification of stochastic disturbances in strongly elastic channel flows. *J. Non-Newtonian Fluid Mech.* **166**, 755- (2011).

11. G. Hariharan, M. R. Jovanovich, S. Kumar, Amplification of localized body forces in channel flows of viscoelastic fluids. *J. Non-Newtonian Fluid Mech.* **260**, 40- (2018).

12. J. Page, T. A. Zaki, Streak evolution in viscoelastic Couette flow. *J. Fluid Mech.* **742**, 520- (2014).

13. D. Bonn, F. Ingremeau, Y. Amarouchene, H. Kellay, Large velocity fluctuations in small-Reynolds-number pipe flow of polymer solutions. *Phys. Rev. E* **84**, 045301 (2011).

14. L. Pan, A. Morozov, C. Wagner, P. E. Arratia, Nonlinear elastic instability in channel flows at low Reynolds numbers. *Phys. Rev. Lett.* **110**, 174502 (2013).

15. B. Qin, P. E. Arratia, Characterizing elastic turbulence in channel flows at low Reynolds number. *Phys. Rev. Fluids* **2**, 083302 (2017).

16. B. Qin, P. F. Salipante, S. D. Hudson, P. E. Arratia, Flow resistance and structures in viscoelastic channel flows at low Re. *Phys. Rev. Lett.* **123**, 194501 (2019).

17. M. Nagata, Three-dimensional finite-amplitude solutions in plane Couette flow: bifurcation from infinity. *J. Fluid Mech.* **217**, 519-27 (1990).

18. R. B. Bird, C. F. Curtiss, R. C. Armstrong, O. Hasager, Dynamics of Polymer Liquids, (Wiley, 1987) vol.1&2.

19. W. Schoppa, F. Hussain. Coherent structure generation in near-wall turbulence. J. Fluid Mech. 453, 57-108 (2002).

20. P. Alfredsson, M. Matsubara. Freestream turbulence, streaky structures and transition in boundary layer flows. *AIAA Pap.* 2000-2334 (2000).

21. A. Varshney, V. Steinberg, Mixing layer instability and vorticity amplification in a creeping viscoelastic flow. *Phys. Rev. Fluids* **3**, 103303 (2018).

22. A. Varshney, V. Steinberg, Elastic Alfven Waves in elastic turbulence. *Nature Communications* **10**, 652 (2019).

23. A. Varshney, V. Steinberg, Drag enhancement and drag reduction in viscoelastic flow. *Phys. Rev. Fluids* **3**, 103302 (2018).

24. Y. Jun, V. Steinberg, Power and pressure fluctuations in elastic turbulence over a wide range of polymer concentrations. *Phys. Rev. Lett.* **102**, 124503 (2009).

25. L. D. Landau and E. M. Lifschitz. *Fluid Mechanics,* (Oxford, Pergamon Press, 1987).

26. W. Thielicke, E. Stamhuis, PIVlab-Towards user-friendly, affordable and accurate digital particle image velocimetry in MATLAB. *J. Open Res. Soft.* **2**, e30 (2014).

27. J. P. Monty, *PhD thesis*, University of Melbourne (2005).

**Acknowledgments:** We thank G. Falkovich and V. Kumar for helpful comments. **Funding:** This work was partially supported by the grants from Israel Science Foundation (ISF; grant #882/15 and grant #784/19) and Binational USA-Israel Foundation (BSF; grant #2016145). PBC Fellowship at Weizmann institute of Science for NKJ is gratefully acknowledged. **Author contributions.** V.S. conceived the project and together with N.K.J. planned the experiments. N.K.J. performed the measurements and analyzed the data. V.S. wrote the manuscript, and both the authors discussed the results and commented on the manuscript. **Competing interests:** Authors declare no competing interests. **Data and materials availability:** All data in the main text or the supplementary materials is available on request.

**Supplementary Materials** for the paper is available at…

Supplementary text

Figures S1-S10

Movie S1-S3

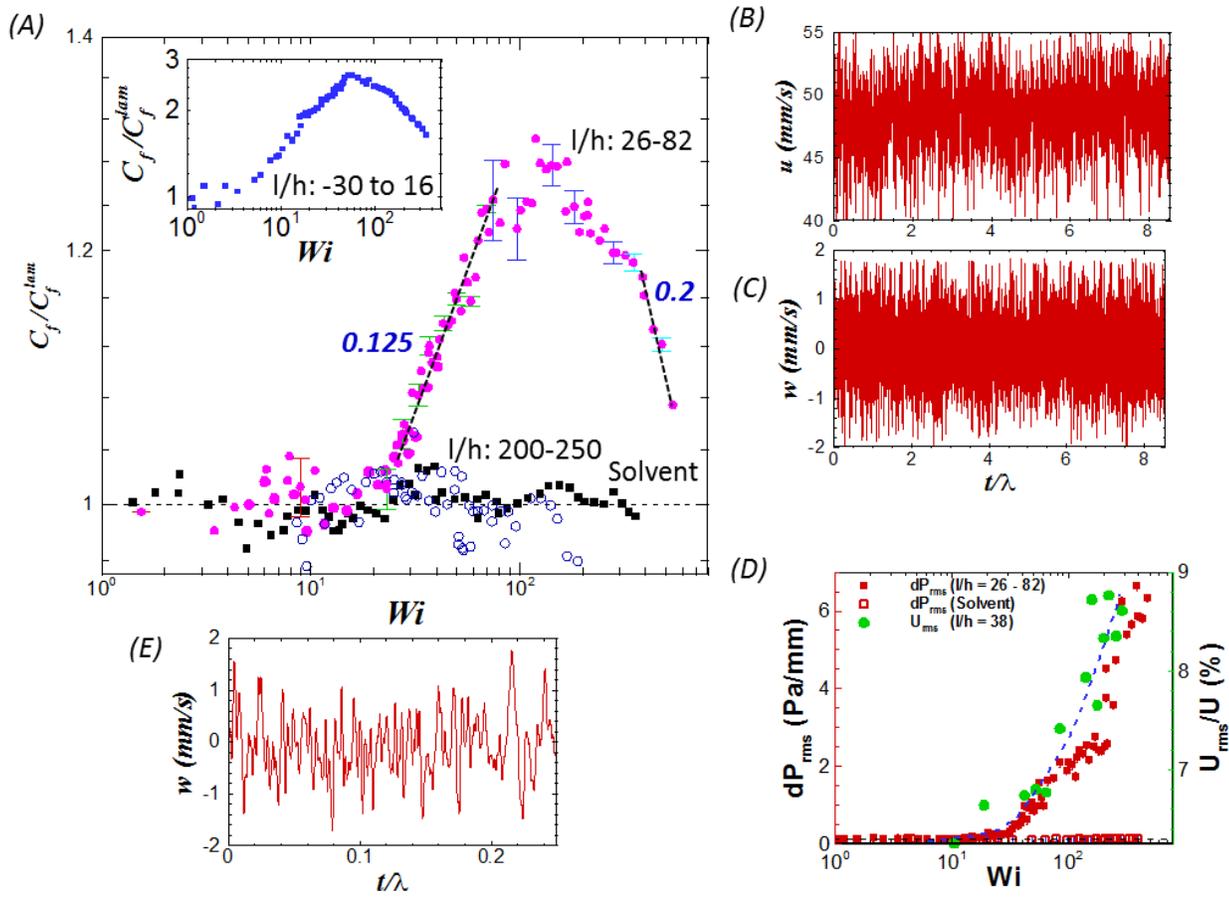

**Fig. 1. Effect of elastic turbulence on friction factor, velocity and pressure.** (A) Dependence of the normalized friction factor $C_f/C_f^{lam}$ on Wi for three values of non-dimensional distance from the third layer of cylinders (at l/h=0) l/h: (i) -30 to 16 (inset), (ii) 26 to 82, and (iii) 200 to 250. Inset shows $C_f/C_f^{lam}$ versus Wi across three rows of cylinders. (B) and (C) show strong fluctuations of stream-wise (u) and span-wise (w) velocities, respectively, in ET at Wi=185 and l/h=38.6. (D) The dependence of the normalized rms pressure difference $dP_{rms}$ (for l/h = 26-82) and normalized rms stream-wise velocities, $u_{rms}/U(\%)$ on Wi at l/h = 38.6. A clear growth is noticed for all variables above the transition. (E) Fluctuations of *w* at Wi=185 and l/h=38.6 on short time scale.

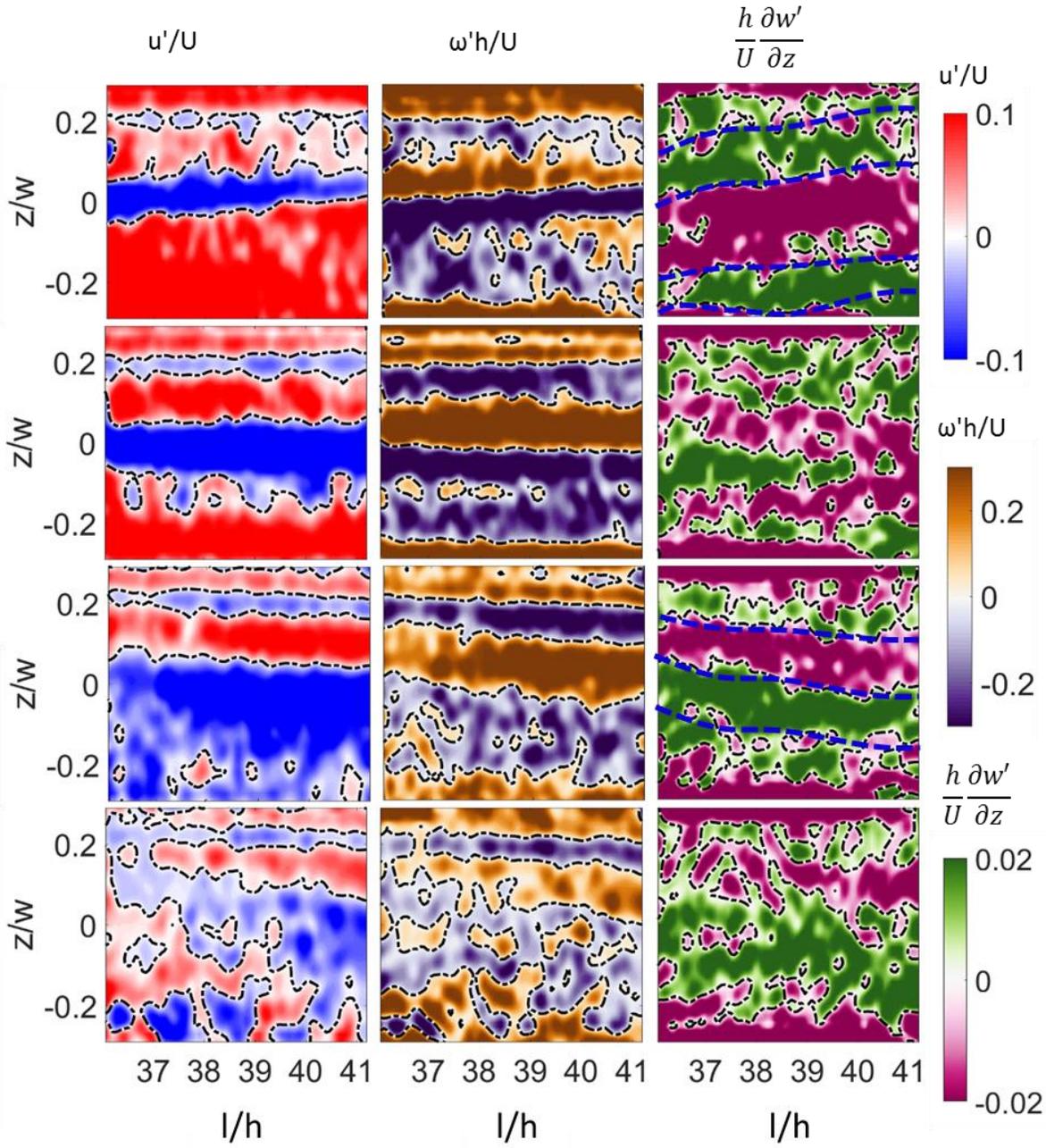

**Fig. 2. A cycle of coherent structures in ET at Wi=185 and l/h = 36-41.** The normalized time t*=tf$_{el}$, where f$_{el}$ is the elastic wave frequency. Fluctuating stream-wise velocity (u'), vertical vorticity (ω'), and span-wise gradient of span-wise velocity (∂w'/∂z) are shown in each column with their scales shown on the right as marked in the plot. Flow structures at four t* in a cycle are shown in four rows and are for t* of 0.15, 0.29, 0.56, and 0.67 starting from top. At t* = 0.15, coherent structure (CS) of stream-wise streaks of high and low speed are shown as positive and negative stream-wise fluctuating velocity (u') separated by dotted black lines of u'=0, presenting also span-wise perturbations. In ω' plot, strips of cross-stream vorticity at the center of streaks is identified together with its random pattern near the edge of strips. Contour of ∂w'/∂z presents stream-wise vortex (roll), another CS, together with a random structure. At t*= 0.29, CSs of u' and ω' look even more pronounced, whereas ∂w'/∂z field is significantly perturbed. At t* = 0.56, streaks and cross-stream vorticity CS are strongly perturbed, but the roll CS is more pronounced in ∂w'/∂z, marked by dotted blue lines and this time all CSs co-exist. Finally, at t* = 0.67, all fields become chaotic, and at later time, a new cycle starts. Full movie of the cycle of coherent structures can be seen in movie S1.

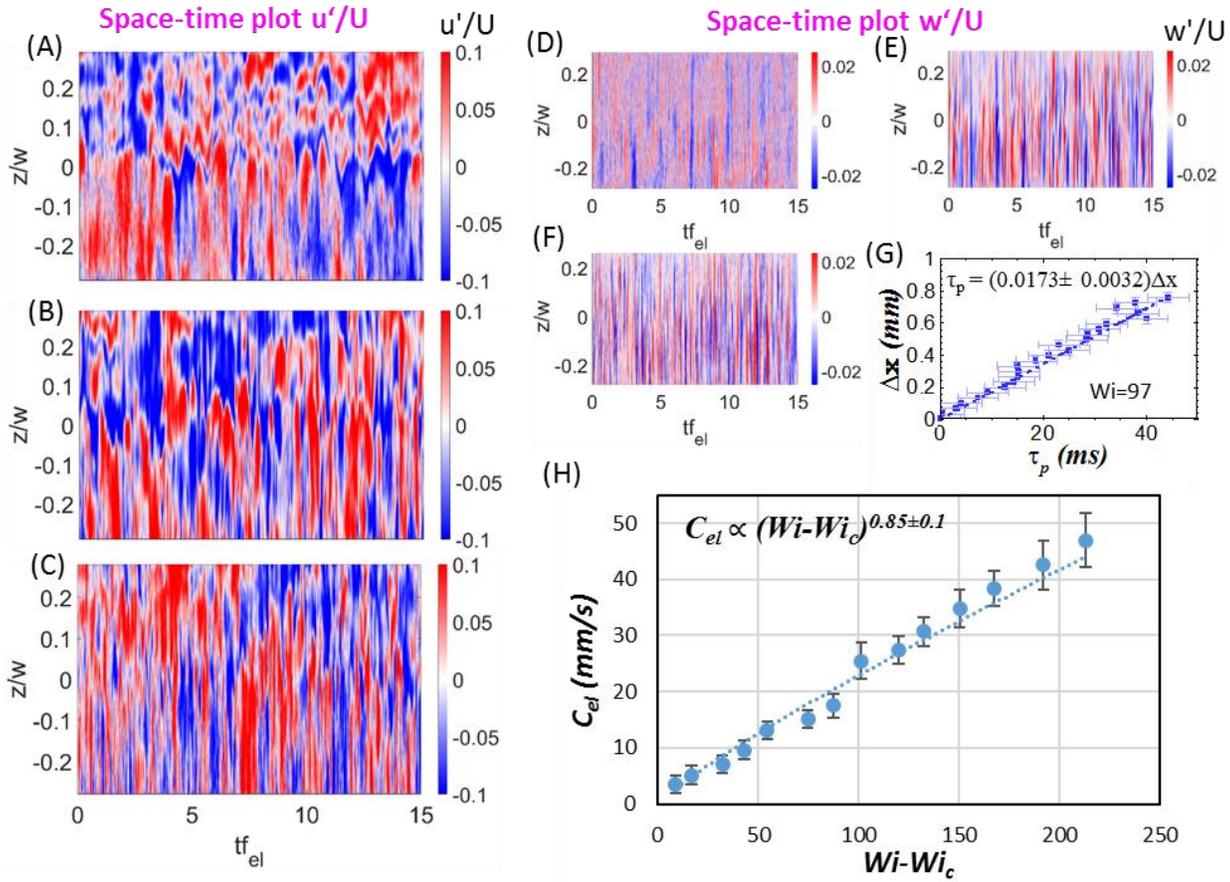

**Fig 3:** **Space-time plot** of u' (A) Wi = 55; (B) Wi = 185; (C) Wi = 325 and w' (D) Wi = 55; (E) Wi = 185; (F) Wi = 325 at l/h of 38.6 in span-wise plane (z) along with the characterization of elastic wave in figure (G) and (H). Three values of Wi = 55, 185, 325 are chosen to cover three different regimes (transition, ET and DR) in $C_f$ /$C_l^{lam}$ versus Wi dependence. Time axis in (A) to (F) is normalized by $f_{el}$, z by the channel width $w$, and $u'$ and $w'$ by the mean stream-wise velocity (U). Middle of the channel corresponds to z=0. Space-time plot of $u'$ shows a reduction of structure scale with increasing Wi, and noisy waviness can be identified in w ' plot, which increases with Wi. Figure (G) shows the shift in location of temporal cross-correlation peak for different stream-wise distance for Wi=97. Slope of the plot is the elastic wave speed $c_{el}$, which dependence on Wi is presented in figure (H). The power-law dependence $c_{el}$~(Wi–$Wi_c$)^ς has the exponent ς=0.85±0.1.

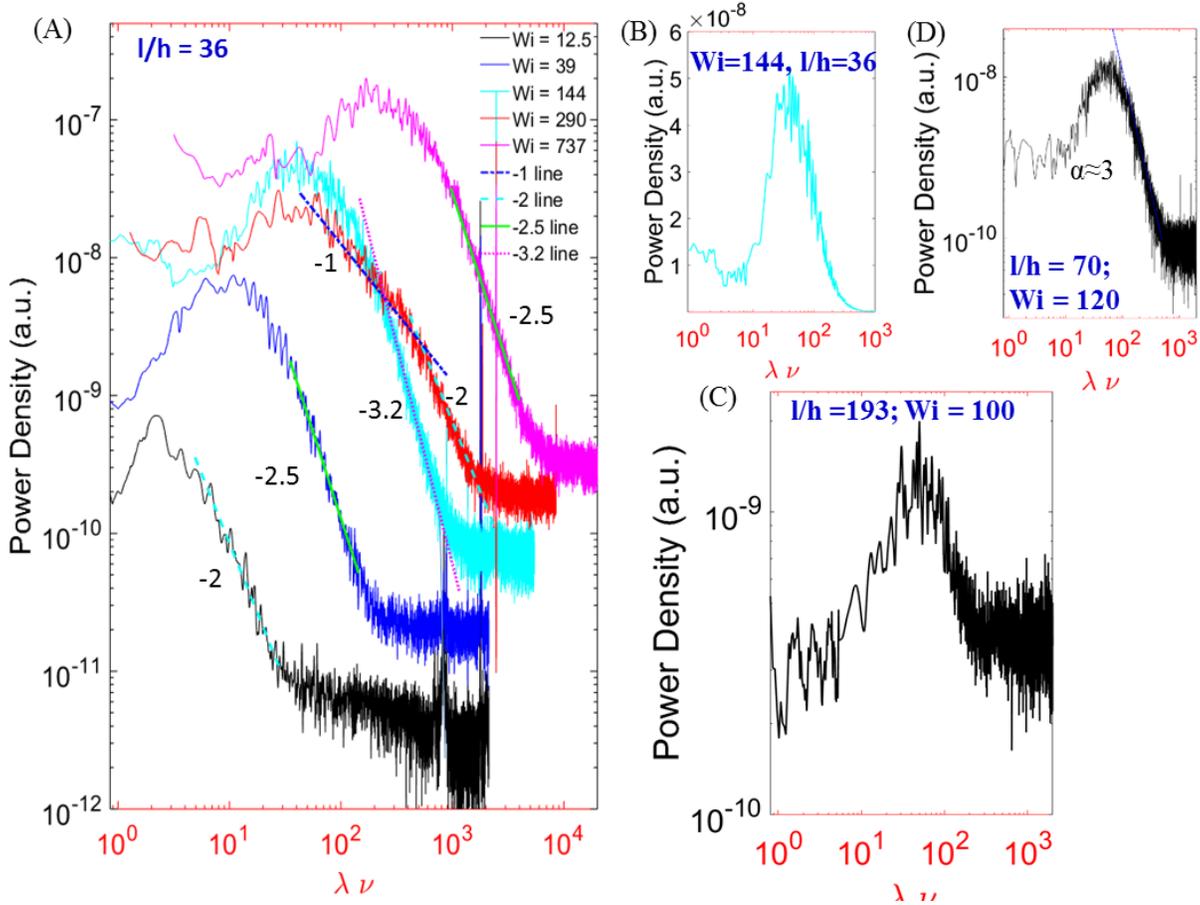

**Fig 4. Velocity power spectra.** (A) Span-wise velocity *w* power spectra for various Wi at l/h=38.6. As already discussed in *(21)*, all span-wise velocity spectra exhibit a wide peak at low frequency, which looks very pronounced in lin-log scale (B). At larger frequencies all spectra at l/h=38.6 show the power-law decay in *(A)* with the exponents depending on the flow regime from $\alpha$=2 and 2.5 in the transition region up to 3.2 in ET, back to 2 in the DR regime and then again rises to 2.5 in post DR at even much higher Wi. (C) The low frequency peak is observed even at l/h=193, where it is found just next to velocity spectrum decay with $\alpha \approx$ 1 at Wi=120. The low frequency peak indicates the elastic waves. As pointed out in *(21)*, the elastic waves pump energy from the large to small length scales in the decay range of the velocity power spectrum (D).

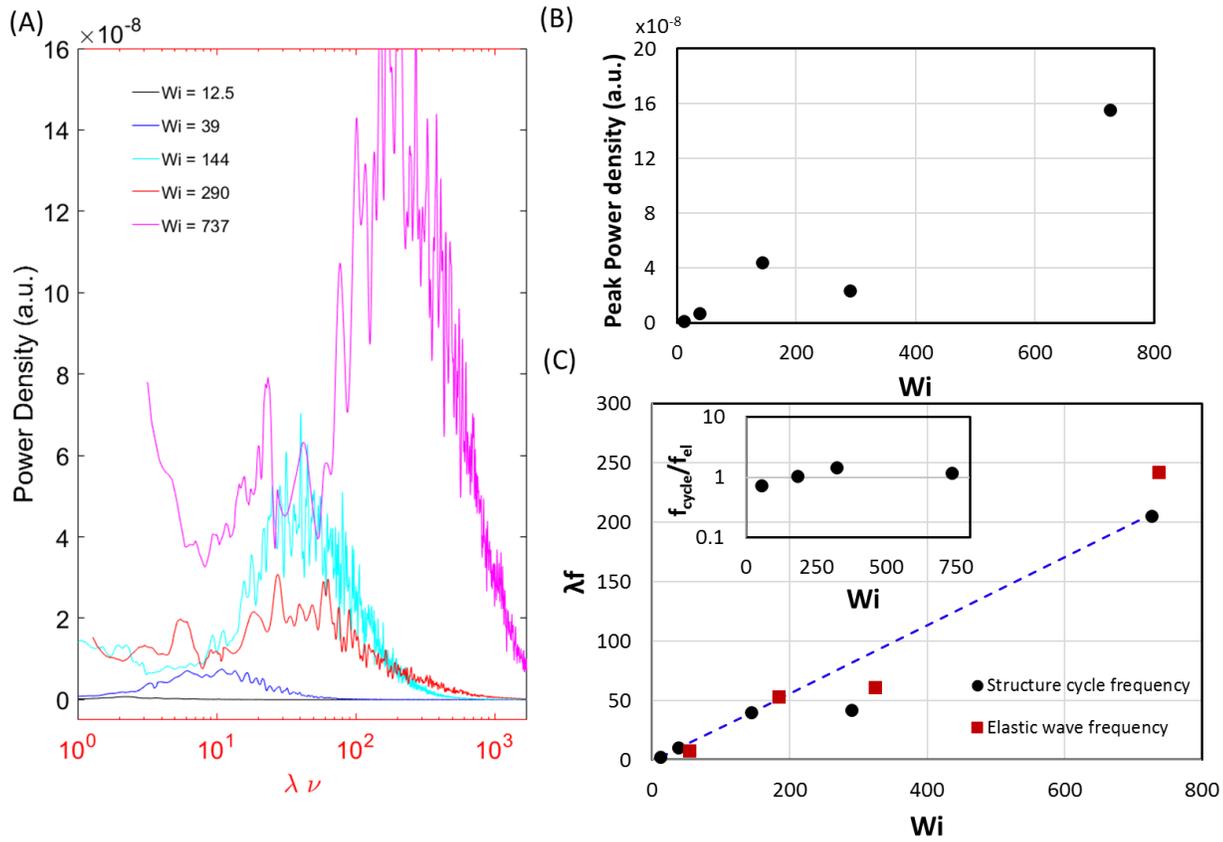

**Fig. 5. (A) Span-wise velocity *w* power spectra in lin-log scale** to compare the maximum values of energy peaks at different Wi in different flow regimes at l/h=38.6 are presented. (B) Plot of the peak values in the velocity power spectra vs Wi shows monotonic increase except of a drop for DR regime. (C) Dependence of structure cycle frequency ($f_{cycle}$) and elastic wave frequency ($f_{el}$) on Wi at l/h=36-41. Inset: Dependence of $f_{cycle}/f_{el}$ on Wi at l/h=36-41.

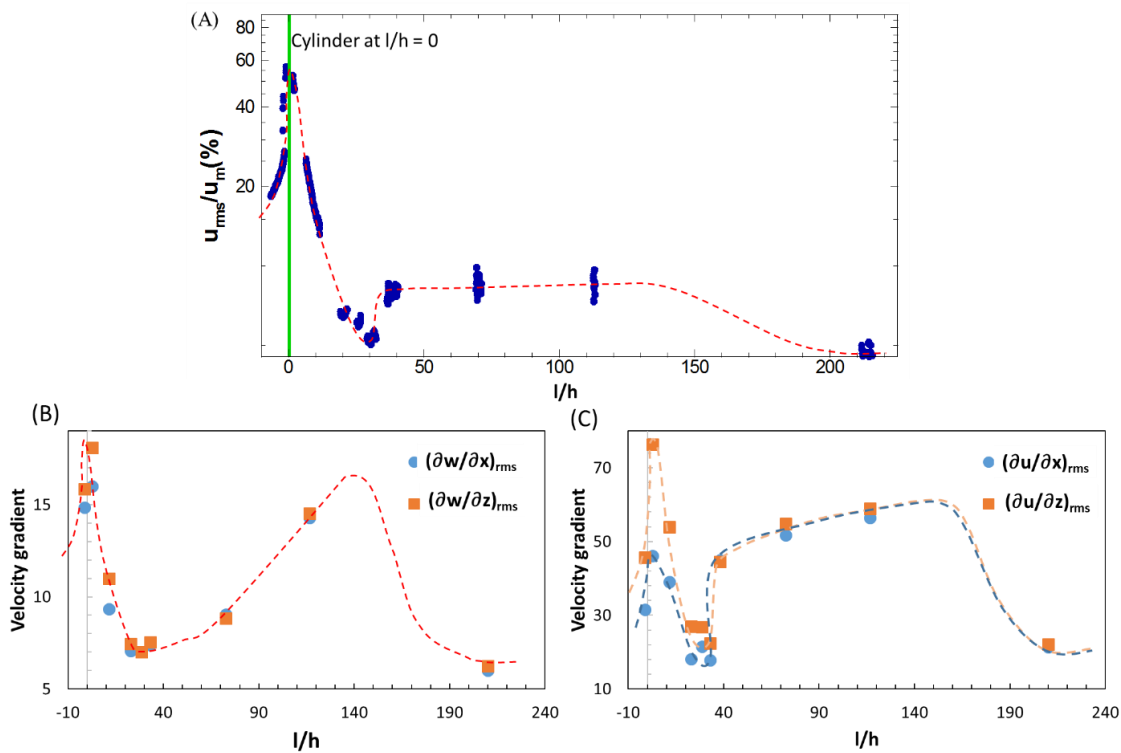

**Fig. 6. (A) Dependence of $u_{rms}/u_m$ on l/h for Wi=150, (B) and (C) Dependence of rms velocity gradients on l/h for Wi=150, which is a representation of elastic stress.** Surprising similarity in spatial distribution of both the span-wise and stream-wise velocity gradients observed in ET regime indicates that elastic stress and polymer stretching is isotropic, though the flow field is highly anisotropic. However, near the cylinders the velocity gradients $\partial u'/\partial x$ and $\partial u'/\partial z$ in x and z directions are anisotropic being higher in z direction similar to *(15)*. Here $u_m$ is the local mean velocity. Curves with dashed lines are shown to highlight the trend of the data and also to demarcate among different variables.

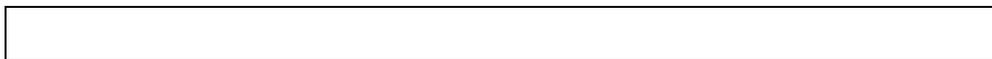

**Supplementary Materials for**

**Universal coherent structures of elastic turbulence in straight channel with viscoelastic fluid flow**

Narsing K. Jha[1] and Victor Steinberg[1,2*]

*Correspondence to: victor.steinberg@weizmann.ac.il

## Supplementary Notes

*Additional description of the results presented in Fig. S2.* First, we comment on the schematic drawing that illustrates the way to detect a stream-wise vortex/roll via measurement of $\frac{\partial w'}{\partial z}$ in xz-plane (Fig. S2). We use this approach, since we are not able to conduct measurements in vertical plane using only 2D PIV in xz-plane. In order to detect a stream-wise vortex one needs to shift PIV-plane out of the vortex center. Otherwise, a projection of the rotation velocity vector would have zero component on the PIV-plane. Thus, the PIV-plane is located the distance d apart of the central plane. An angle α between the R, the vortex radius, and d is defined by Cosα=d/R. Moving the point of intersection of R with d along the PIV-plane towards the vortex central vertical line (in a span-wise direction z), the corresponding $\alpha_i$ is varied and so is defined by $Cos\alpha_i=d/R_i$. It is easy to show that for the vortex rotating as a solid body $V_i=\Omega R_i$, where $V_i$ is the azimuthal velocity vector at location $z_i$, one gets $w_i=\Omega d$=constant and $\frac{\partial w'}{\partial z}$=0. But, in a general case of a rotational flow $V_i=\Omega f(R_i)$ and $w_i=\Omega f(R_i)d/R_i$ one finds $\frac{\partial w'}{\partial z}\neq0$ due to dependence $R_i(z_i)$. Thus the velocity gradient $\frac{\partial w'}{\partial z}$ in the stream-wise vortex can be detected in x-z plane as seen in Fig 2, S3, S4, S5.

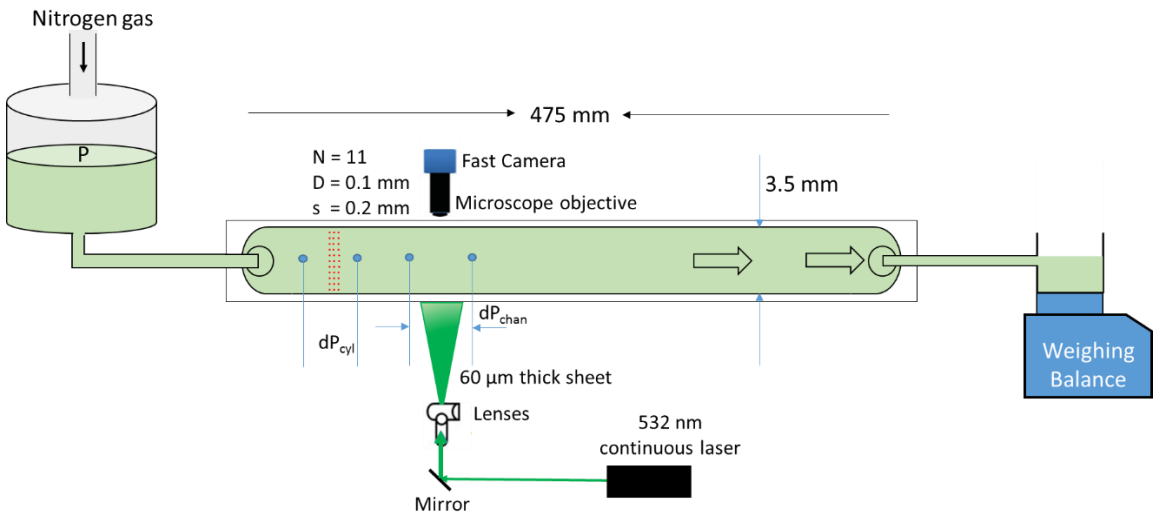

**Fig.S1.** ***Schematic of experimental setup***. Straight channel of dimensions $l \times w \times h = 475 \times 3.5 \times 0.5 \ mm^3$ made of PMMA has the array of cylindrical obstacles of 100µm diameter organized in 3 rows with 11 cylinders in each row separated by 200µm both stream- and span-wise. Such cylinder array produces a rather strong perturbation across the full channel width, in contrast to a square channel used in *(14-16)*. Our straight channel has quasi-2D rectangular geometry, and its length ratio l/h=950 is about 3 times larger and aspect ratio w/h=7 is 7 times larger than that of the square channel used in *(14-16)*. Two pressure sensors are used to measure the pressure drop across the obstacles array $\Delta P_{cyl}$ and in the channel $\Delta P_{chan}$ over the appropriate distance $\Delta L$, and pressure fluctuations are detected by separate absolute pressure sensor. The polymer solution is driven by the Nitrogen gas pressurized up to 100 psi. To obtain the friction factor the fluid average velocity <U>=Q/ρwh is calculated via the average fluid discharge Q=<Δm/Δt>, where m(t) is weighed instantaneously as a function of time. Illuminated by a laser sheet, velocity field is measured via PIV.

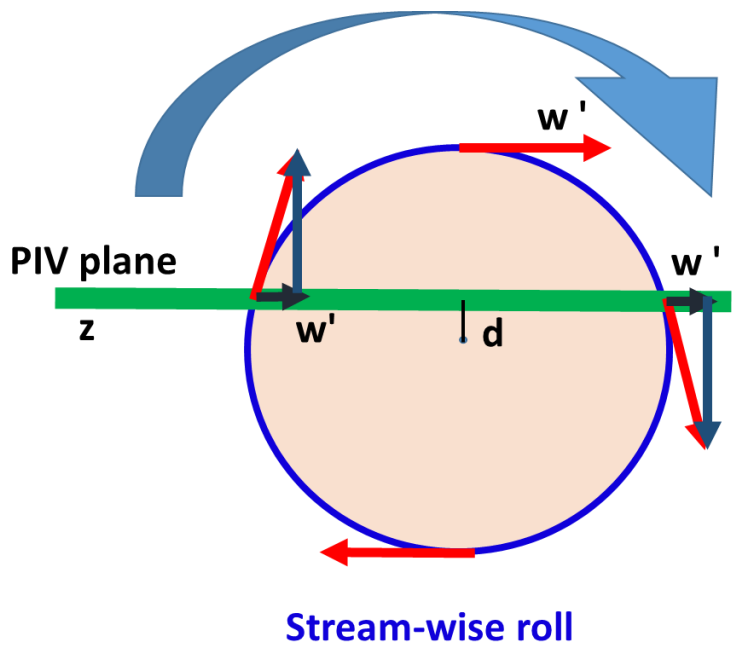

**Fig. S2. Detection of stream-wise vortex from 2D PIV in a horizontal xz plane.**

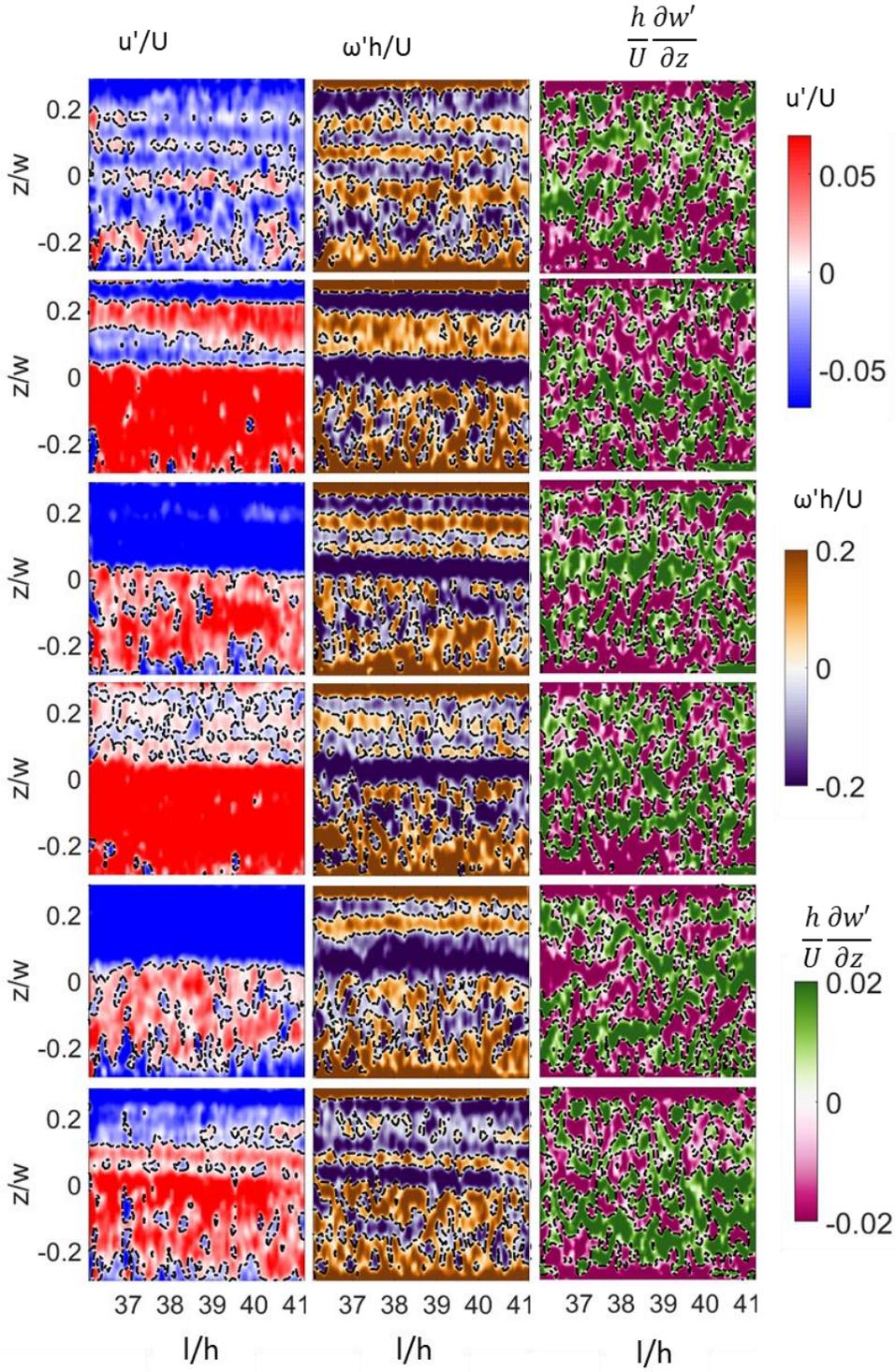

**Fig. S3. Spatial flow structures in a transitional regime at Wi=55 and l/h=36-41 during a cycle.** The normalized time $t^*=tf_{el}$, where $f_{el}$ is the elastic wave frequency. $t^*$ increases from top to bottom and is constant for each row. Stream-wise velocity (u'), vertical vorticity (ω'), and span-wise gradient of span-wise velocity ($\frac{\partial w'}{\partial z}$) are marked above each column with their scales shown on the right. Flow structures at six $t^*$ are shown in six rows and present a cycle. At $t^* = 0.04$, almost completely random structures for all three variables with very narrow and irregular stream-wise streaks of fluctuating velocity u' and cross-stream vorticity ω' are seen. Further at $t^* = 0.25$, the high and low speed streaks of u' and ω' start appearing. At next three values of $t^*$ at 0.35, 0.46, and 0.62, the streaks of u' as well as ω' are separated by rather sharp interfaces marked by dotted black lines of u'=0, together with random flow structure of $\frac{\partial w'}{\partial z}$. And for the last t*=1.2 the streaks in both u' and ω' are broken but the breakdown is not complete and clear. Thus at Wi=55 in the transition regime, only the streaks are found, the stream-wise vortices (rolls) are absent, and cycling period is observed. Full movie of the cycle of coherent structures can be seen in movie S2.

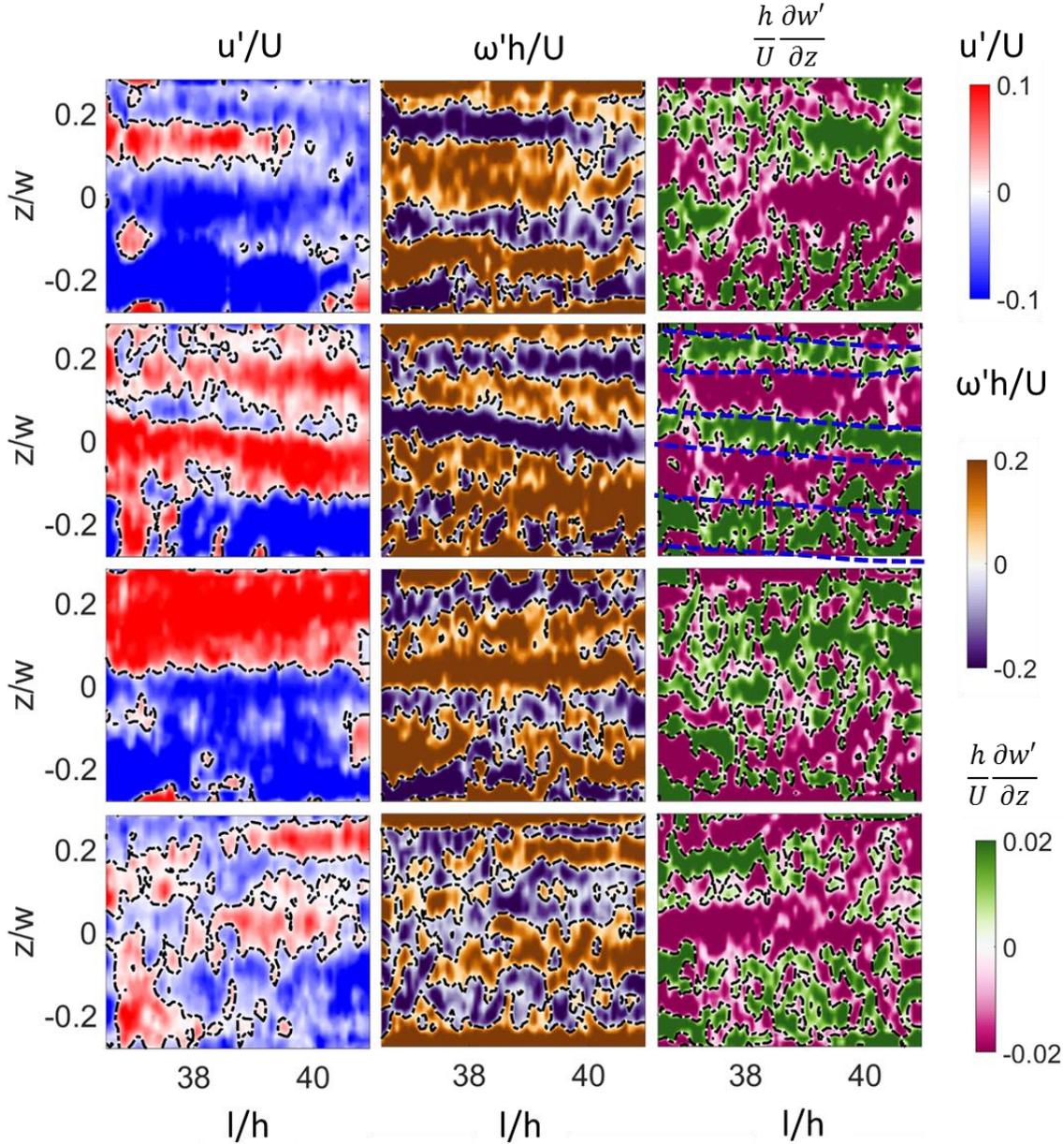

**Fig. S4. Spatial flow structures in DR regime at Wi = 325 and l/h=36-41 during a cycle.** As in Fig. S3, the normalized stream-wise velocity (u'), vertical vorticity (ω'), and span-wise gradient of span-wise velocity ($\partial w'/\partial z$) are marked above each column with their scales shown on the right, and t* increases from top to bottom and is constant for each row. Flow structures at four t* values in a cycle are shown in four rows. At t* = 0.23 and t*=0.45, CS of stream-wise streaks of high and low speed are shown as positive and negative stream-wise fluctuating velocity u' separated by dotted black lines of u'=0, presenting also span-wise perturbations. In ω' plot, streaks of the cross-stream vorticity ω' are observed, though they are not well identified, similar to streaks, due to strong perturbations. Rolls of $\partial w'/\partial z$ are also not clearly defined due to perturbations at t*=0.23 and at t*=0.45 a random structure is found, whereas at t*=0.3 $\partial w'/\partial z$ field presents a single stream-wise

vortex pair together with a random structure. At t*= 0.3, CSs of u' and ω' show a streak pair. Finally at t* = 0.69, all fields become mostly chaotic, and at later time, a new cycle starts. Full movie of the cycle of coherent structures can be seen in movie S3.

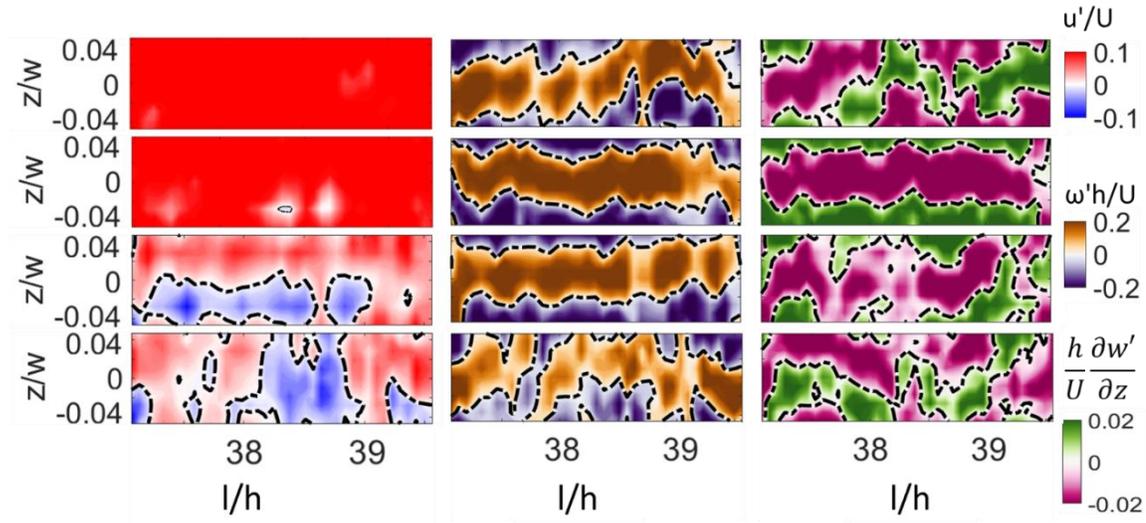

**Fig. S5. Spatial flow structures in ET regime at Wi=737 and l/h=37-39.8 during a cycle.** Images are narrower than at other Wi due to higher U and so frame rate that leads to lower spatial resolution of full velocity field. At such high Wi, CSs, both streaks and rolls, are defined much clearly than for lower Wi values, and a cycle repeats more precisely. At t* of 0.09, high speed streak is observed, and as the time passes, at t* of 0.2, high speed streak is observed along with streamwise vortex. At t* of 0.35, both high and low speed highly perturbed streaks appear and streamwise vortex is destroyed by perturbations. Finally at t* of 0.78, CSs are broken down and becomes random, and a new cycle starts. t* increases from top to bottom and is constant for each row.

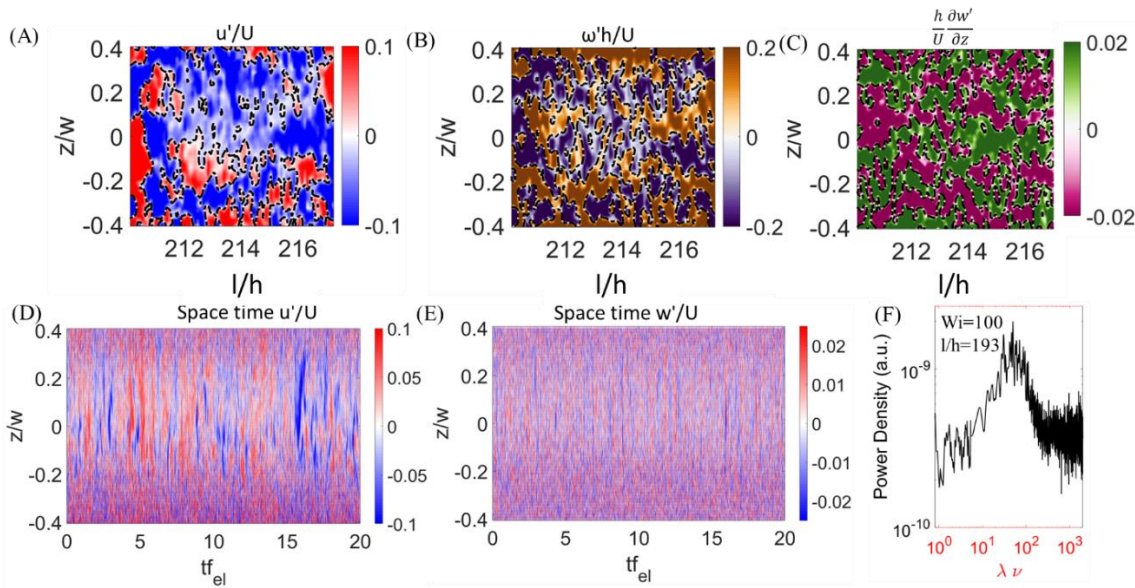

**Fig. S6. Spatial flow structures and u/U and w/U̲ space-time plots with the corresponding velocity _w_ power spectrum at Wi=146 and l/h = 210 − 217.** All flow structures are random (A-C). The space-time plots (D-E) shows noisy periodicity presented by a wide and noisy peak in the span-wise velocity w power spectrum of very low intensity. As the result, the characteristic for ET the power spectrum decays is absent.

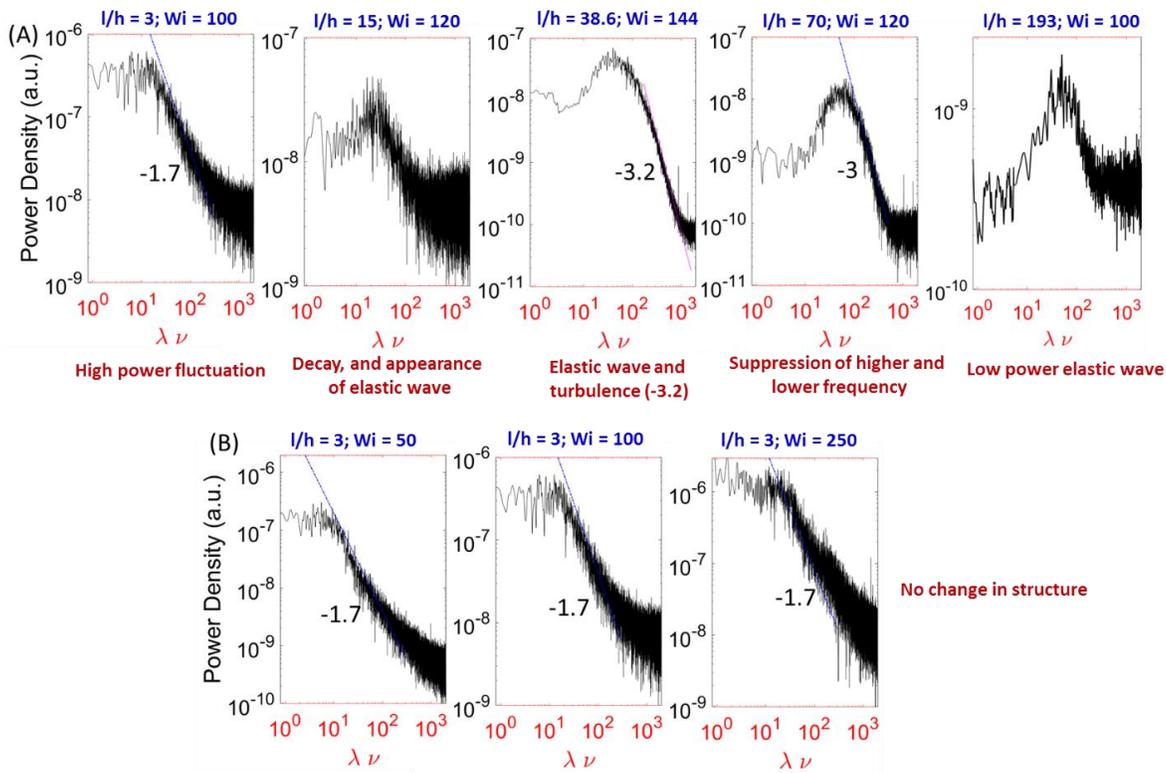

**Fig. S7. Span-wise velocity *w* power spectra at l/h=3, 15, 38.6, 70 and 193 at fixed Wi values.** (A) Examples of the kinetic energy spectra at different l/h and Wi values in the the ET regime. (B) Three kinetic energy spectra at l/h=3 and wide range of Wi values show similar shape with increasing intensity.

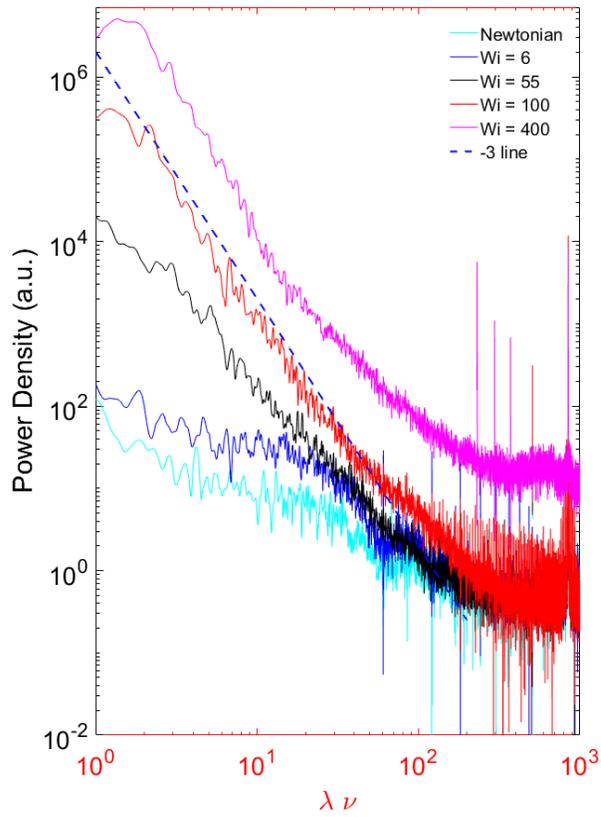

**Fig. S8. Temporal pressure spectra at different Wi in three flow regimes (transitional, ET, DR).**

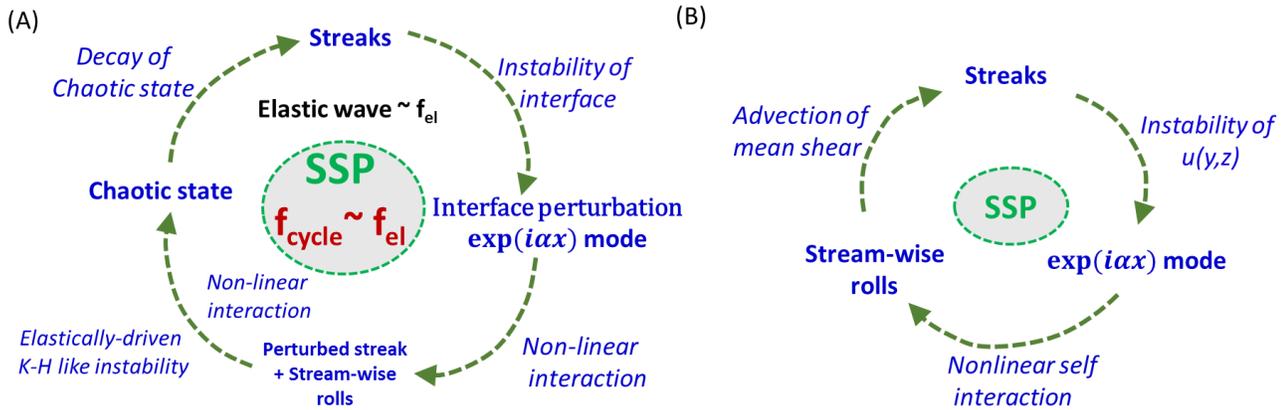

**Fig. S9. Schematic of CS cycling in viscoelastic (A) compared with Newtonian fluid (B) channel flows.** (B) is adopted from *(1)*. Numerical simulations of *(11,12)* are unable to capture the nonlinear interaction part and SSP cycle presented in (A).

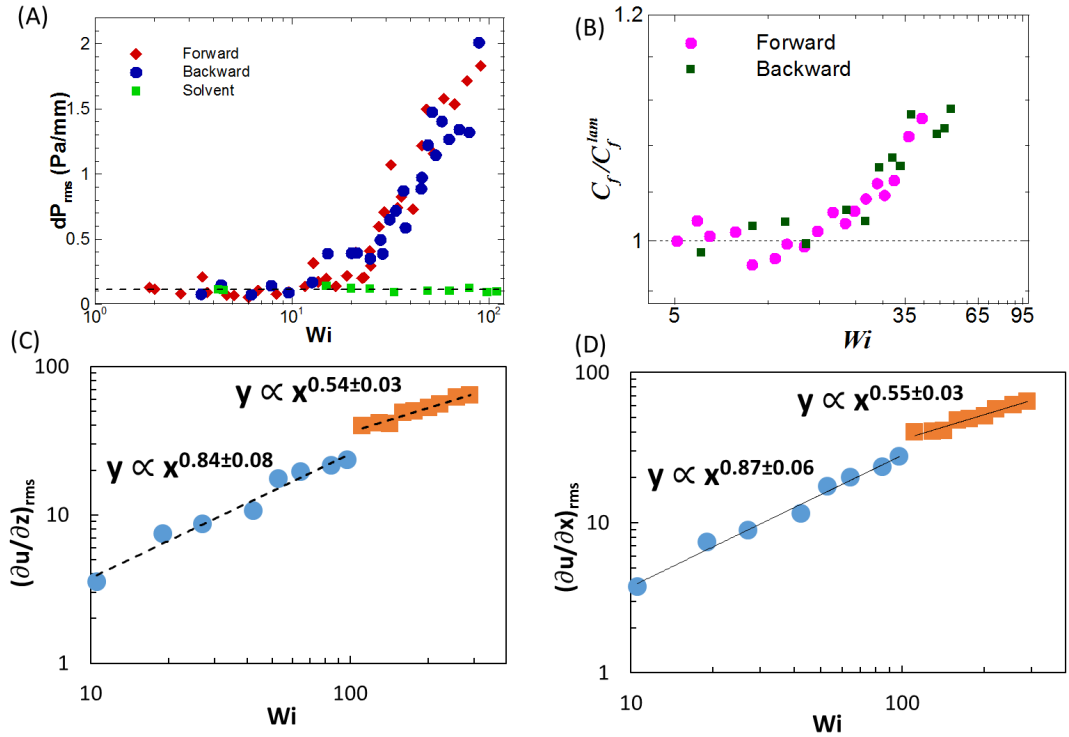

**Fig. S10. (A) Test of hysteresis in the dependence of dP$_{rms}$ on Wi at l/h=26-82. (B) Test of hysteresis in C$_f$/C$_f^{lam}$ at l/h=26-82.** Dependence of (C) $(\partial u'/\partial z)_{rms}$ and (D) $(\partial u'/\partial x)_{rms}$ on Wi for l/h=38.6. Surprising similarity in both the rms gradients of the stream-wise velocity in all flow regimes is found indicating that elastic stress and polymer stretching is isotropic even when the velocity field is highly anisotropic. Exponents are higher before transition and reduces to about 0.5 after transition to ET.

**Movie S1. Spatial flow structure in ET at Wi=185 and l/h = 36-41 during a cycle.** The normalized time t*=tf$_{el}$, where f$_{el}$ is the elastic wave frequency. Normalized stream-wise velocity (u'), vertical vorticity (ω'), and span-wise gradient of span-wise velocity ($\frac{\partial w'}{\partial z}$) are marked above each window with their scales shown either below or on the right.

**Movie S2. Spatial flow structure in a transitional region at Wi=55 and l/h=36-41 during a cycle.** The normalized time t*=tf$_{el}$, where f$_{el}$ is the elastic wave frequency. Stream-wise velocity (u'), vertical vorticity (ω'), and span-wise gradient of span-wise velocity ($\frac{\partial w'}{\partial z}$) are marked above each column with their scales shown either below or on the right.

**Movie S3. Spatial flow structure in DR regime at Wi = 325 and l/h=36-41 during a cycle.** As in **Movie S2**, normalized stream-wise velocity (u'), vertical vorticity (ω'), and span-wise gradient of span-wise velocity ($\partial w'/\partial z$) are marked above each column with their scales shown either below or on the right.